\documentclass[letterpaper]{article}
\usepackage{aaai}
\usepackage{times}
\usepackage{helvet}
\usepackage{graphicx}
\usepackage{nameref}
\usepackage{courier}
\usepackage{amsmath}
\usepackage{booktabs}
\usepackage{multirow}
\usepackage{float}
\usepackage{tabularx}
\restylefloat{table}
\usepackage{siunitx} 
\begin{document}
%
\title{Fetal Gender Identification using Machine and Deep Learning Algorithms on Phonocardiogram Signals}
\author{Reza Khanmohammadi \textsuperscript{\rm 1}, Mitra Sadat Mirshafiee \textsuperscript{\rm 2}, Mohammad M. Ghassemi \textsuperscript{\rm 3*}, Tuka Alhanai \textsuperscript{\rm 4*}\\
Department of Computer Engineering, University of Guilan, Iran \textsuperscript{\rm 1}\\
Department of Industrial Engineering, Alzahra University, Iran \textsuperscript{\rm 2}\\
Department of Computer Science, Michigan State University, USA \textsuperscript{\rm 3}\\
Department of Electrical \& Computer Engineering, New York Univesity Abu Dhabi, UAE \textsuperscript{\rm 4}\\
rezanecessary@gmail.com, mitra.mirshafiee@gmail.com, ghassem3@msu.edu, tuka.alhanai@nyu.edu\\
}
\maketitle
\begin{abstract}
\begin{quote}
Phonocardiogram (PCG) signal analysis is a critical, widely-studied technology to noninvasively analyze the heart's mechanical activity. Through evaluating heart sounds, this technology has been chiefly leveraged as a preliminary solution to automatically diagnose Cardiovascular diseases among adults; however, prenatal tasks such as fetal gender identification have been relatively less studied using fetal Phonocardiography (FPCG). In this work, we apply common PCG signal processing techniques on the gender-tagged Shiraz University Fetal Heart Sounds Database and study the applicability of previously proposed features in classifying fetal gender using both Machine Learning and Deep Learning models. Even though PCG data acquisition's cost-effectiveness and feasibility make it a convenient method of Fetal Heart Rate (FHR) monitoring, the contaminated nature of PCG signals with the noise of various types makes it a challenging modality. To address this problem, we experimented with both static and adaptive noise reduction techniques such as Low-pass filtering, Denoising Autoencoders, and Source Separators. We apply a wide range of previously proposed classifiers to our dataset and propose a novel ensemble method of Fetal Gender Identification (FGI). Our method substantially outperformed the baseline and reached up to 91\% accuracy in classifying fetal gender of unseen subjects. 
\end{quote}
\end{abstract}

\section{Introduction}
Gender detection has been extensively investigated in computer science.  There is a large body of research on the use of images \cite{Matkowski_2020}, text \cite{Yildiz}, and time-series data \cite{Gornale} to study this problem, as well as a range of technologies to classify phenomena according to gender \cite{SWAMINATHAN20202634,9180161}. However, prenatal gender identification of fetuses \cite{Ornoy}, an essential clinical procedure, has received relatively less attention mainly due to data acquisition difficulty and certain additional challenges that pregnancy poses \cite{Sameni}.

Meanwhile, in a totally different field of research, abundant studies are examining ways to monitor heart activity to detect Cardiovascular Diseases (CVDs) using either classical Machine Learning (ML) or Deep Learning (DL) algorithms \cite{Suyi,e23060667}. Heart diseases are responsible for 38\% of all premature deaths globally \cite{who}, thus developments in ML-based heart monitoring would yield a significant improvements in preventative health. One practical technological approach for monitoring the heart is Phonocardiography, from which Phonocardiogram (PCG) signals are attained. These signals contain a sound representation of the heart's mechanical activity, for which signal acquisition does not rely on expensive medical equipment, which is usually the case. 

\subsection{Contributions}
In this work, we take a step towards automatic Fetal Gender Identification (FGI) with an understudied approach of using Fetal Phonocardiogram (FPCG) signals. More specifically, we experiment with multiple ML and DL pipelines containing a variety of preprocessing, feature extraction, and classification techniques to map fetal cardio patterns to the gender of the fetus. However, PCG signals suffer from various noise and require certain denoising techniques to be initially applied for signals to become gender-detectable. We train a Denoising Autoencoder and a neural source separation technique to measure the effects of two fundamentally different audio cleansing techniques. Afterwards, we evaluate top PCG classification methodologies in CVD detection on our gender-tagged FPCG dataset to establish a firm FGI baseline. Finally, we propose a novel ensemble method that outperforms the rest of the classifiers in FGI. We use medically accepted metrics such as accuracy, sensitivity, and specificity \cite{Trevethan} to evaluate the efficiency of each method in the hold-out and leave one out settings.

The remaining of this paper is organized as follows: first, we describe previous studies in the \nameref{sec:Related} section. Then we fully describe our approach in the \nameref{sec:Methodology} section and compare the result with previous work in \nameref{sec:Results} before concluding the paper in the last section.

\section{Related Work} \label{sec:Related}
\subsection{Advantages of Phonocardiography Signals}
\noindent Automatic classification of heartbeat signals has attracted considerable interest due to the practicality and effectiveness of proposed approaches in detecting Cardiovascular Diseases (CVDs) \cite{Ismail}. Generally, the principal difference between the majority of such research is the monitoring method with which the heartbeats have been collected. Among noninvasive methods, Electrocardiography (ECG) has been relatively more studied than expensive methods such as Magnetocardiography (MCG) or less accurate methods such as Photoplethysmography (PPG) \cite{Avanzato}. Because of the timing and sequence of the heart's depolarization and repolarization, ECG signals can be collected by putting electrodes on different parts of the body surface to detect the electrical activity of different portions of the heart. Such signals consist of three major waves, which are recalled as \textit{P} wave, \textit{QRS} complex, and \textit{T} wave, which respectively represent the impact of atrial depolarization, ventricular depolarization, and ventricular repolarization in the final ECG signal. Still, extracting the FHR using this method has shown to be rather time-consuming and difficult to apply on pregnant women \cite{Sameni}. In contrast, Phonocardiography signals, which record the mechanical activity of the heart during cardiac cycles as murmurs and sounds, are known to be an entirely passive, cost-effective alternative. When obtained from adults, these waveforms contain two fundamentally normal sounds (namely, S1 and S2) and two other potentially abnormal sounds (S3 and S4). During development, the fetal heart differs from the adult heart, where the Patent Doramen Ovale (PDO) allows the blood flow to bypass the lungs \cite{PDOC}, which ends up FPCG signals having their unique characteristics. However, such a dissimilarity is not the only challenge that is posed in FPCG analysis.

\subsection{Noise Processing Techniques}
A certain impediment to adopting PCG signals in computational tasks is the necessity to handle various low and high-frequency noise through preprocessing \cite{Yang2020CLASSIFICATIONOP}. As \cite{Tseng2021CrossDomainTL} suggests, noise  generally originates either from the recording devices (instrumental noise) or the surroundings (ambient noise). The first can be a result of the microphone\textquotesingle{s} internal electronics\textquotesingle{} functionality; however, its adverse effects are relatively minor in an analogy to the latter where a broad range of sources can lower signal quality. Background noise, muscle contractions, and respiratory sounds \cite{Suyi} are noises are challenging for researchers to process and derive meaningful information from PCG signals \cite{LEAL2018154,Pham}. Since signal quality heavily influences the derivation of features \cite{Suyi}, researchers have investigated denoising techniques to come up with proper filters that differentiate heartbeat signals from the unwanted \cite{Tseng2021CrossDomainTL}. Common PCG denoising filters are either static such as Butterworth low-pass filtering \cite{Gomes2012ClassifyingHS}, or adaptive such as the least mean square algorithm \cite{Dewangan2014NoiseCU}. Empirical Mode Decomposition (EMD), Intrinsic Mode Functions (IMFs), and Nonnegative Matrix Factorization (NMF) are among other techniques that have been successfully employed to individually/jointly clean a given signal  \cite{Pham}. Other solutions leverage source separators \cite{Shiyu}, or autoencoder networks \cite{Chuang2019SpeakerAwareDD}, to distinguish heartbeat sounds - the latter technique is one we will further investigate in the \nameref{sec:Methodology} section of this paper. 

\subsection{Featurizing Phonocardiography Signals}
The success of PCG signal classification heavily depends on advancements in machine learning \cite{Suyi}. Classifiers such as Support Vector Machines (SVM) \cite{8714943} or k-Nearest Neighbours (KNN) \cite{Juniati} have previously succeeded in detecting abnormalities such as murmurs or extrasystole. However, these advancements heavily rely on handcrafted discriminant features that get fed as inputs to such models \cite{LIJUNIII}. In our case, PCG signals are characteristically two-folded: (1) By adopting a signal processing perspective, studies have investigated features such as mean, variance, or power in each of the time \cite{DENG201613}, frequency \cite{SAFARA20131407}, and time-frequency domains \cite{THIYAGARAJA2018313}. (2) Whereas another point to process these recordings is to capture their sound features. Mel Spectrogram \cite{9183915}, and Mel-frequency Cepstrum (MFCC) \cite{app8122344} are among the most widely leveraged feature representations techniques defining sound characteristics using a specifiable number of coefficients.

Depending on the problem of interest, conditions regarding different PCG datasets, and applied preprocessing techniques, studies have focused on each individually or combined the two to train classifiers \cite{Kilin}. Based on previous results and achievements, there is greater confidence that CVDs can almost accurately be detected using state-of-the-art ML-based methodologies.

\subsection{Deep Learning}
In the meantime, DL algorithms have followed the same trend and are no exceptions to this sense of confidence \cite{e23060667}. Standing on Neural Networks\textquotesingle{} superiority where the extraction and selection of features are eradicated to be individually stepped through, plenty of studies have experimented with Multilayer Perceptron (MLP) \cite{s19214819}, Convolutional Neural Networks (CNN) \cite{WU201929}, and Recurrent Neural Networks (RNN) \cite{latif2020phonocardiographic}. Unlike ML algorithms, no handcrafted features are fed to these networks and instead, low-level sets of features are extracted to represent the signals. MFCC \cite{Jonathan}, Spectrogram \cite{Demir}, or Constant-Q transform \cite{9508967} are among the most frequently used sound feature representations that have substantially aided related studies. However, the favorability of their outcomes has heavily been dependent on the abundance of informative data \cite{khan2021deep}. An interesting research line has recently tackled PCG signal classification using CNNs on the image represented sound features \cite{Ren2018LearningIR}. In this setting, low-level sound representations are converted into images, and CNNs are asked to extract image features to classify heartbeats. Widely employed in image processing tasks, pretrained image networks such as VGGnet \cite{simonyan2015deep} and ResNet \cite{he2015deep} have also been leveraged to improve classification accuracy. Research on CVD detection using Transfer Learning \cite{9287514} has allowed researchers to more effectively handle smaller datasets where scarce data cannot be fed to neural networks directly, thereby unfavourably improving the chances of overfitting \cite{9115633}.

\section{Methodology} \label{sec:Methodology}
To classify fetal gender using FPCG signals, we use the version 1.0.1 of the publically available Shiraz University Fetal Heart Sounds Database \cite{Sameni}. This database contains PCG signals of 109 pregnant women and FPCG signals recorded by placing a JABEZ\textsuperscript{TM} electronic stethoscope on their lower maternal abdomen. There were 7 cases of pregnancies where the mother was carrying a twin, which we excluded in this work, resulting in 102 samples which further are broken down to 7-second samples, among which a balanced number of male and female signals are considered to form our final dataset of 1000 FPCG samples. The original sampling rate of these recordings is 16,000 Hz and are, on average, 90 seconds long. A large number of previous studies have experimented with PCG samples of 5 to 10 seconds using major datasets (e.g. PhysioNet \cite{Qiao} or PASCAL \cite{Coimbra}). To balance between the informativeness of chunked samples and their final count, we chose seven seconds as the maximum length of each sound sample. To do so, we used the peak detection algorithm of the SciPy Python library \cite{2020SciPy} to find positive (heartbeats) and negative (silence) peaks from sound files and thus avoid cutting a signal when the heart beat cycle was not complete. To further describe our work, hereafter, we step through a pipeline to denoise, classify, and stack models for better efficiency.

\begin{figure}[!t]
\label{fig:3in1}
    \centering
    \includegraphics[width=\linewidth]{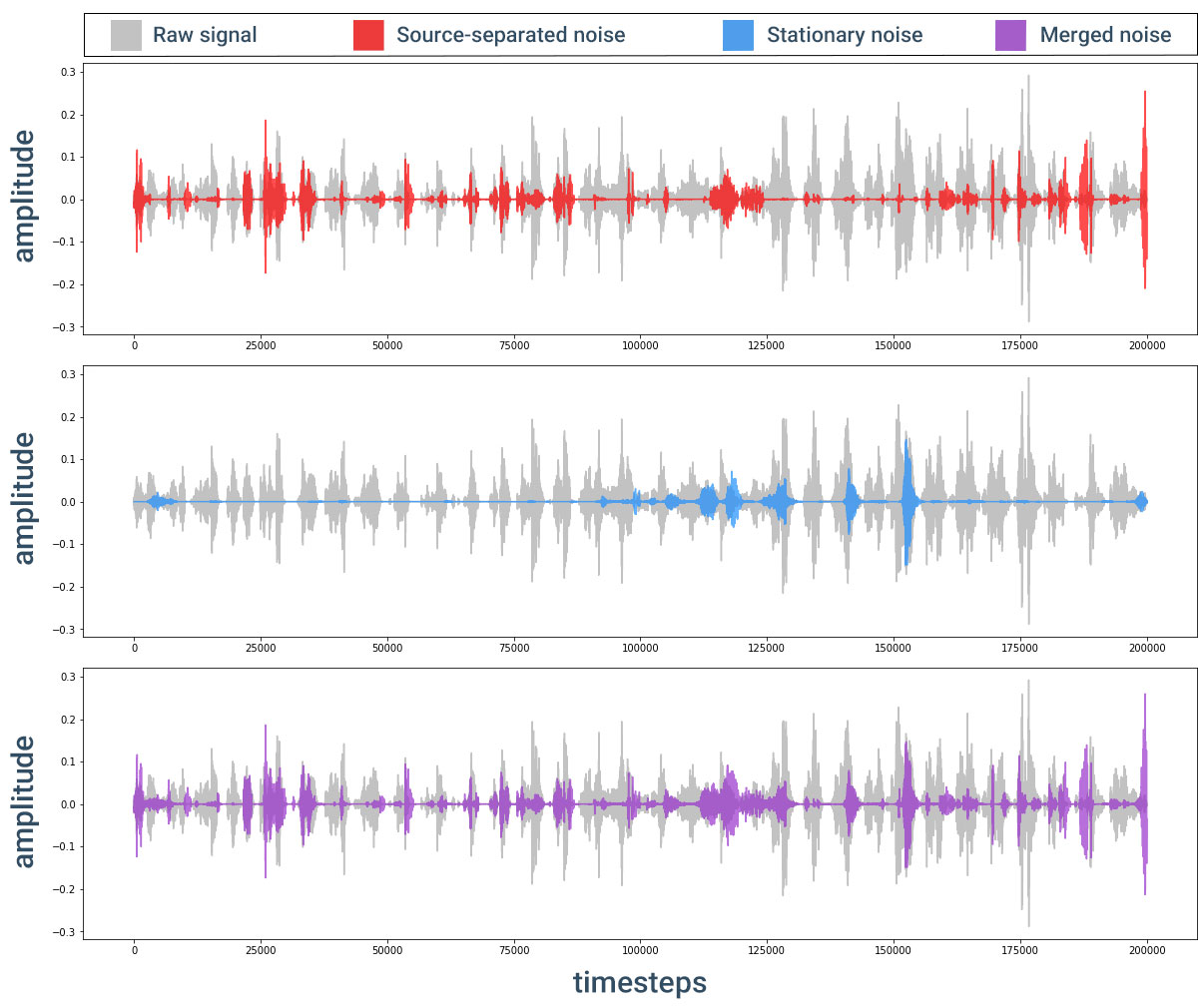}
    \caption{An example of how Single-Channel Blind Source Seperation (SCBSS) using a deep learning framework (top), and Stationary filtering (middle) may help us detect noise among Phonocardiography (FPCG) signals. By merging the two, a final noise waveform (bottom) is obtained. Original waveform is plotted in gray, while detected noise is colored.}
    \label{fig:my_label1}
\end{figure}

\subsection{Denoising}
Removing noise is of great importance since our dataset's FPCG signals contain samples among which heartbeats may either have a high or low frequency. This condition also reigns in terms of noise, where the low-frequency (e.g. the friction between the stethoscope's surface and human skin) and high-frequency (e.g. background noise) noise are diversely distributed. Hence, in our case, simply low-pass filtering \cite{s19214819} FPCG signals using static thresholds cannot distinguish heartbeats from noise. Instead, what we need is an adaptive filter that intelligently differentiates the two. Hence, we experiment with the following well-studied techniques to do so.

\subsubsection{Denoising Autoencoders (DAE):} 
An autoencoder consists of an encoder (\(e)\) and a decoder (\(d)\) that are connected through an encoded state. The encoder learns to map an input vector \(x(i)\) into a compressed mid-level representation \(z(i)\) such that the decoder is better modelled to reconstruct output vector \(y(i)\) from \(z(i)\). 
\begin{equation}
\label{eq:1}
    z[i] = e(x[i])
\end{equation} 
\begin{equation}
\label{eq:2}
    y[i] = d(z[i])
\end{equation}

Given two parallel sets of noisy and clean signals, we can train an autoencoder network with the objective to denoise signals. Here we use the PASCAL dataset, which contains plenty of clean PCG signals and common clinical artifacts. This way, through combining a random artifact with a clean PCG signal, we obtain a noisy \(x(i)\) from which the autoencoder will learn to generate clean heartbeats as \(y(i)\).

\subsubsection{Single-Channel Blind Source Separation (SCBSS):} we use a deep learning framework called DeWave \cite{hershey2015deep,caron2019deep} to apply SCBSS on FPCG signals using the deep clustering algorithm. To separate audio sources, this method uses embedding vectors for each time-frequency bin. As indicated in equation 3, they have designed their objective function to make embedding vectors \(v_i\) more similar/dissimilar when two bins do/do not belong to the same class.  
\begin{equation}
\label{eq:3}
    \begin{split}
            & C(\theta, Y) =\left|V V^{T}-Y Y^{T}\right|_{\mathrm{F}}^{2}=\\
            & \sum_{i, j: y_{i}=y_{j}}\left(\left|v_{i}-v_{j}\right|^{2}-1\right)+\sum_{i, j} \frac{1}{4}\left(\left|v_{i}-v_{j}\right|^{2}-2\right)^{2}
    \end{split}
\end{equation}
Given \(VV^{T}\) as a low-rank affinity matrix and \(YY^{T}\) as an ideal affinity matrix, network parameters \(\theta\) are trained to minimize the objective function. Whereas the first term compacts bins of the same class, the second term spreads out bins of different classes. To apply this technique, we need to train the network on two distinct audio sources to model certain patterns of each. We again use the PASCAL dataset by feeding its artifact samples and pure heartbeats as two audio sources to the framework. At inference, given a single noisy heartbeat sound signal (\(s_{s}(i)\)), we obtain two audio files named \(s_{h}(i)\) (heartbeat signal) and \(s_{n}(i)\) (noise signal) where each belongs to one of the two audio source. 

\begin{equation}
\label{eq:4}
    s_{h}[i], s_{n}[i] = w(s_{s}[i])
\end{equation} 

Since the DeWave method (\(w\)) was initially proposed and evaluated on speech data, its effects have not been fully investigated in auscultation. However, in terms of our FPCG signals, they perform interestingly well in detecting background, instrumental, and continuous noise, whereas they fail to capture sudden noisy fluctuations. Next, to better detect mishandled high-frequency noise, we use a spectral gating noise reduction technique (\(g\)) \cite{sainburg2020finding}. In this case, we used the stationary variant of this technique since a statically estimated threshold captures our noise of interest much better. Finally, we merge \(s_{n}(i)\) with \(g(s_{s}(i))\) that of the stationary filter to obtain a final noise waveform (\(m(i)\)). 
\begin{equation}
\label{eq:5}
    m[i] = s_{n}[i] + g(s_{s}[i])
\end{equation} 
This way, given a noisy FPCG signal as \(s_{s}(i)\), we obtain \(m(i)\), which is the merged waveform of \(s_{s}(i)\) high and low-frequency noise. To filter a waveform inside another, we take the Fourier transforms (\(F\)) of both \(s_{s}(i)\) and \(m(i)\) to further apply a band-stop filter over certain frequencies of \(F(s_{s}(i))\) where \(F(m(i))\) has high magnitudes in. Finally, by taking the Inverse Fourier Transform of the filtered \(F(s_{s}(i))\), the final denoised signal \(s_{d}(i)\) is created.


\begin{table*}
  \centering
  \footnotesize
  \begin{tabular}{|c|c|c|c|cccc|cccc|}
    \hline
    \multirow{2}{*}{Classifier} &
    \multirow{2}{*}{Input} &
    \multirow{2}{*}{Denoising method} &
    \multirow{2}{*}{Classifier} &
    \multicolumn{4}{c|}{Hold-out} & 
      \multicolumn{4}{c|}{LOOCV} \\
      &  &  &  & Acc & PR & SN & SP & Acc & PR & SN & SP \\
     \hline
     \hline
    \multirow{ 2}{*}{} &  & SCBSS & KNN & 0.59 & 0.38 & 0.76 & 0.52 & 0.58 & 0.37 & 0.74 & 0.52 \\
     &  & . & SVM & 0.6 & 0.49 & 0.9 & 0.43 & 0.59 & 0.45 & 0.94 & 0.40 \\
     &  & . & XGB & 0.56 & 0.47 & 0.83 & 0.38 & 0.53 & 0.51 & 0.68 & 0.39 \\
     &  & . & \textbf{LDA} & \textbf{0.66} & \textbf{0.55} & \textbf{0.8} & \textbf{0.57} & \textbf{0.63} & \textbf{0.47} & \textbf{0.88} & \textbf{0.51} \\
     &  & . & LR & 0.65 & 0.75 & 0.52 & 0.80 & 0.55 & 0.54 & 0.49 & 0.62 \\
     \(C_{R}(x)\) & \(R(x)\) & DAE & KNN & 0.56 & 0.41 & 0.58 & 0.55 & 0.55 & 0.45 & 0.48 & 0.60 \\
     &  & . & SVM & 0.6 & 0.47 & 0.89 & 0.43 & 0.58 & 0.42 & 1.0 & 0.40 \\
     &  & . & XGB & 0.55 & 0.51 & 0.70 & 0.41 & 0.54 & 0.51 & 0.67 & 0.42 \\
     &  & . & LDA & 0.62 & 0.55 & 0.66 & 0.58 & 0.57 & 0.41 & 0.75 & 0.48 \\
     &  & . & LR & 0.50 & 0.45 & 0.56 & 0.45 & 0.47 & 0.31 & 0.87 & 0.34 \\
    \hline
    \multirow{ 2}{*}{} &  & SCBSS & KNN & 0.52 & 0.39 & 0.86 & 0.35 & 0.49 & 0.31 & 0.84 & 0.37 \\
     &  & . & SVM & 0.53 & 0.41 & 0.64 & 0.46 & 0.51 & 0.37 & 0.72 & 0.41 \\
     &  & . & \textbf{XGB} & \textbf{0.66} & \textbf{0.54} & \textbf{0.75} & \textbf{0.61} & \textbf{0.63} & \textbf{0.48} & \textbf{0.78} & \textbf{0.54} \\
     &  & . & LDA & 0.58 & 0.51 & 0.63 & 0.54 & 0.55 & 0.43 & 0.65 & 0.49 \\
     &  & . & LR & 0.54 & 0.46 & 0.68 & 0.44 & 0.50 & 0.54 & 0.47 & 0.54 \\
     \(C_{M}(x)\) & \(M(x)\) & DAE & KNN & 0.59 & 0.35 & 0.90 & 0.49 & 0.63 & 0.47 & 0.72 & 0.57 \\
     &  & . & SVM & 0.58 & 0.40 & 0.95 & 0.43 & 0.57 & 0.43 & 0.83 & 0.43 \\
     &  & . & XGB & 0.65 & 0.58 & 0.72 & 0.60 & 0.61 & 0.49 & 0.75 & 0.52 \\
     &  & . & LDA & 0.61 & 0.46 & 0.80 & 0.5 & 0.52 & 0.37 & 0.76 & 0.41 \\
     &  & . & LR & 0.51 & 0.40 & 0.59 & 0.46 & 0.49 & 0.29 & 0.97 & 0.36 \\
    \hline
    \multirow{ 2}{*}{} &  & SCBSS & KNN & 0.66 & 0.66 & 0.72 & 0.60 & 0.64 & 0.66 & 0.69 & 0.59 \\
     &  & . & \textbf{SVM} & \textbf{0.71} & \textbf{0.58} & \textbf{0.86} & \textbf{0.62} & \textbf{0.68} & \textbf{0.51} & \textbf{0.98} & \textbf{0.53} \\
     &  & . & XGB & 0.67 & 0.58 & 0.77 & 0.60 & 0.65 & 0.61 & 0.70 & 0.61 \\
     &  & . & LDA & 0.59 & 0.51 & 0.72 & 0.49 & 0.58 & 0.51 & 0.68 & 0.5 \\
     &  & . & LR & 0.57 & 0.48 & 0.70 & 0.48 & 0.55 & 0.55 & 0.57 & 0.53 \\
     \(C_{MF}(x)\) & \(MF(x)\) & DAE & KNN & 0.61 & 0.58 & 0.77 & 0.46 & 0.57 & 0.50 & 0.66 & 0.50 \\
     &  & . & SVM & 0.67 & 0.60 & 0.73 & 0.63 & 0.66 & 0.62 & 0.70 & 0.64 \\
     &  & . & XGB & 0.56 & 0.43 & 0.93 & 0.36 & 0.54 & 0.43 & 0.88 & 0.35 \\
     &  & . & LDA & 0.54 & 0.58 & 0.55 & 0.54 & 0.52 & 0.47 & 0.64 & 0.42 \\
     &  & . & LR & 0.65 & 0.75 & 0.52 & 0.80 & 0.55 & 0.54 & 0.49 & 0.62 \\
    \hline
    \multirow{ 2}{*}{} &  & SCBSS & \textbf{KNN} & \textbf{0.67} & \textbf{0.52} & \textbf{0.93} & \textbf{0.53} & \textbf{0.65} & \textbf{0.53} & \textbf{0.86} & \textbf{0.52} \\
     &  & . & SVM & 0.60 & 0.34 & 0.72 & 0.56 & 0.57 & 0.28 & 0.68 & 0.54 \\
     &  & . & XGB & 0.63 & 0.52 & 0.79 & 0.54 & 0.62 & 0.51 & 0.78 & 0.51 \\
     &  & . & LDA & 0.54 & 0.37 & 0.93 & 0.39 & 0.54 & 0.46 & 0.63 & 0.47 \\
     &  & . & LR & 0.59 & 0.40 & 0.58 & 0.59 & 0.57 & 0.30 & 1.0 & 0.47 \\
     \(C_{Q}(x)\) & \(Q(x)\) & DAE & KNN & 0.61 & 0.49 & 0.89 & 0.45 & 0.60 & 0.44 & 0.98 & 0.41 \\
     &  & . & SVM & 0.61 & 0.48 & 0.51 & 0.68 & 0.57 & 0.31 & 0.97 & 0.47 \\
     &  & . & XGB & 0.59 & 0.44 & 0.81 & 0.47 & 0.56 & 0.50 & 0.62 & 0.51 \\
     &  & . & LDA & 0.52 & 0.37 & 0.87 & 0.37 & 0.52 & 0.49 & 0.55 & 0.5 \\
     &  & . & LR & 0.52 & 0.4 & 0.49 & 0.54 & 0.53 & 0.41 & 0.53 & 0.53 \\
    \hline
    \multirow{ 2}{*}{} &  & SCBSS & KNN & 0.54 & 0.58 & 0.53 & 0.57 & 0.51 & 0.50 & 0.54 & 0.49 \\
     &  & . & SVM & 0.68 & 0.57 & 0.81 & 0.6 & 0.75 & 0.70 & 0.79 & 0.71 \\
     &  & . & \textbf{XGB} & \textbf{0.91} & \textbf{0.86} & \textbf{0.96} & \textbf{0.86} & \textbf{0.89} & \textbf{0.87} & \textbf{0.91} & \textbf{0.87} \\
     &  & . & LDA & 0.72 & 0.65 & 0.82 & 0.64 & 0.7 & 0.67 & 0.75 & 0.65 \\
     & \(T_{1-7}(x)\), & . & LR & 0.56 & 0.41 & 0.85 & 0.42 & 0.53 & 0.49 & 0.59 & 0.48 \\
     \(C_{S}(x)\) & \(F_{1-35}(x)\), & DAE & KNN & 0.56 & 0.55 & 0.6 & 0.52 & 0.59 & 0.49 & 0.74 & 0.48 \\
     & \(TF_{1-28}(x)\) & . & SVM & 0.87 & 0.79 & 0.96 & 0.80 & 0.89 & 0.83 & 0.95 & 0.84 \\
     &  & . & XGB & 0.86 & 0.78 & 0.96 & 0.78 & 0.83 & 0.73 & 0.97 & 0.73 \\
     &  & . & LDA & 0.71 & 0.59 & 0.68 & 0.72 & 0.68 & 0.51 & 0.72 & 0.66 \\
     &  & . & LR & 0.63 & 0.45 & 0.75 & 0.58 & 0.71 & 0.63 & 0.75 & 0.68 \\
    \hline
    \multirow{ 2}{*}{} &  & SCBSS & KNN & 0.81 & 0.79 & 0.82 & 0.79 & 0.76 & 0.61 & 0.96 & 0.64 \\
     &  & . & SVM & 0.91 & 0.93 & 0.92 & 0.90 & 0.82 & 0.72 & 0.91 & 0.77 \\
     &  & . & \textbf{XGB} & \textbf{0.93} & \textbf{0.89} & \textbf{0.96} & \textbf{0.90} & \textbf{0.91} & \textbf{0.86} & \textbf{0.96} & \textbf{0.86} \\
     & \(C'_{R}(x)\), & . & LDA & 0.78 & 0.74 & 0.83 & 0.73 & 0.81 & 0.81 & 0.84 & 0.79 \\
     & \(C'_{M}(x)\), & . & LR & 0.79 & 0.82 & 0.78 & 0.80 & 0.86 & 0.82 & 0.94 & 0.78 \\
     Ensemble & \(C'_{MF}(x)\), & DAE & KNN & 0.75 & 0.68 & 0.84 & 0.68 & 0.80 & 0.73 & 0.90 & 0.71 \\
     & \(C'_{Q}(x)\), & . & SVM & 0.83 & 0.80 & 0.87 & 0.81 & 0.72 & 0.65 & 0.71 & 0.73 \\
     & \(C'_{S}(x)\) & . & XGB & 0.79 & 0.72 & 0.8 & 0.77 & 0.74 & 0.63 & 0.77 & 0.72 \\
     &  & . & LDA & 0.72 & 0.58 & 0.81 & 0.67 & 0.71 & 0.61 & 0.72 & 0.70 \\
     &  & . & LR & 0.80 & 0.74 & 0.86 & 0.76 & 0.71 & 0.64 & 0.68 & 0.73 \\
    \hline
    
  \end{tabular}
  \caption{\label{tab:ourtabel_1}A comparison between the efficiencies of individual models and how each impacts our final Ensemble method. Best results are \textbf{bolded}. Results reported are Accuracy (Acc), Precision (PR), in addition to clinically-used Sensitivity (SN) and Specificity (SP) scores, with validation performed on a Hold-out set and leave-one-out cross-validation (LOOCV).}
\end{table*}  

\subsection{Feature Extraction}
Both the success and failure of ML models are heavily influenced by the quality of features that we extract from our data. The intuition behind extracting handcrafted features from FPCG signals is to convert a \(s_d[i]\) signal of size \((7_{Seconds} \times 16,000_{SamplingRate} = 112,000_{TimeSteps})\) into a much smaller feature vector where certain characteristics are represented numerically. This way, the classification model is fit to distinguish feature vectors of different classes and thus learn to classify FPCG signals by the gender of the fetus. Generally, related work has categorized PCG features as either being statistical or sound features. In this work, as shown below, we represent signals with a set of 12 features, among which 1 to 7 are statistical features and 8 to 12 are sound features.
\begin{enumerate}
    \item \textbf{Mean:} A general understanding of a signal can be grasped by measuring the mean of amplitudes in an audio signal. letting \(x\) be a denoised signal and \(N\) as the total number of its time steps, mean is calculating as below:
    \begin{equation}
    \label{eq:6}
        \bar{x} = \frac{\sum_{i=1}^{N} x_{i}}{N}
    \end{equation}
    \item \textbf{Variance:} Given \(\bar{x}\) as the mean of the signal and \(N\) as the number of time steps, this feature demonstrates how spread-out a signal is. 
    \begin{equation}
    \label{eq:7}
        variance = \sigma^{2}=\frac{\sum(x-\bar{x})^{2}}{N-1}
    \end{equation}
    \item \textbf{Skewness:} As \cite{Tang2018PCGCU} suggests, this feature assumes that noise has a different probability distribution than heartbeats. 
    \begin{equation}
        \label{eq:8}
        Skewness = \frac{3(\bar{x}-\tilde{x})}{\sigma}
    \end{equation}
    Given \(\bar{x}\) as the mean, \(\tilde{x}\) as the median, and \(\sigma\) as the standard deviation of a signal, skewness is calculated with equation \ref{eq:8}.
    \item \textbf{Kurtosis:} Determines how "tailed" the probability distribution is. 
        \begin{equation}
            \label{eq:9}
            kurtosis = \frac{\sum_{i=1}^{N}\left(x_{i}-\bar{x}\right)^{4}}{(N-1) s^{4}}
        \end{equation}  
        Given \(\bar{x}\) as the mean of distribution and \(N\) as the sample's number of observations (timesteps), kurtosis (equation \ref{eq:9}) is quantified where large values stand for the infrequency of the signal.
    \item \textbf{Spectral Entropy:} Entropy is a measurement of how peaky and disorganized a signal is \cite{Toh}. When extracted from PCG signals, this feature also assumes heartbeat signals having low-frequencies whereas high-frequency noise are uniformly distributed . Where \(P(\omega_{i})\) stands for the spectrum of a signal\textquotesingle{s}, the Probability Density Function (equation \ref{10}) is calculated with which the Power Spectral entropy can be now calculated (equation \ref{eq:11}).
    \begin{equation}
        \label{eq:10}
        p_{i}=\frac{\frac{1}{N}\left|P\left(\omega_{i}\right)\right|^{2}}{\sum_{i} \frac{1}{N}\left|P\left(\omega_{i}\right)\right|^{2}}
    \end{equation}
    \begin{equation}
        \label{eq:11}
        PSE = -\sum_{i=1}^{n} p_{i} \ln p_{i}
    \end{equation}
    \item \textbf{Energy:} The Energy feature (equation \ref{eq:12}) captures sudden changes in mechanical energy of the signal \cite{Kudriavtsev}.
    \begin{equation}
    \label{eq:12}
        E = \sum_{n=-\infty}^{\infty}|x[i]|^{2}
    \end{equation}
    \item \textbf{Root Mean Square (RMS):} RMS is used to measure continuous energy, which is obtained through a square root operation of the mean of the audio signal squared. 
    \begin{equation}
    \label{eq:13}
        RMS = \sqrt{\frac{\sum_{i=1}^{n} x_{i}^{2}}{N}}
    \end{equation}
    \item \textbf{zero-crossing rate (ZCR):} As a popular acoustic feature in speech processing, ZCR informs us of the number of times the signal crosses the horizontal axis.
    \begin{equation}
    \label{eq:14}
    \footnotesize
        ZCR_{t}=\frac{1}{2} \times \sum_{k=t \cdot K}^{(t+1) \cdot K-1}|\operatorname{sgn}(x(k))-\operatorname{sgn}(x(k+1))|
    \end{equation}
    \item \textbf{Chroma (\(R(x))\):} Majorly used to measure the tonal content of audio pieces \cite{Kattel}. 
    \item \textbf{Mel Spectrogram (\(M(x))\):} As a variant of spectrograms, this technique generates perceptually relevant amplitude/frequency representations.
    \item \textbf{Mel-frequency cepstrum (\(MF(x))\):} MFCCs have widely been used as audio feature representations of the short-term power spectrum of a sound.
    \item \textbf{Constant-Q Transform (\(Q(x))\):} CQT simply transforms data from the time domain to the time-frequency domain with large Q-factors \cite{Schrkhuber2010CONSTANTQTT}.
\end{enumerate}

Our goal is to improve our understanding of fetal heartbeats by transforming an FPCG signal from the time domain to the frequency and time-frequency domains. This way, a signal's characteristic gets unveiled in two more domains with which we can better model FGI. Below is a description of how feature extraction is applied in each domain:
\begin{itemize}
    \item \textbf{Time domain (12 features):} Features 1 to 7 \(T_{1-7}(x)\), Chroma mean \(\bar{R}(i)\), Mel Spectrogram mean \(\bar{M}(i)\), MFCC mean \(\bar{MF}(i)\), and CQT mean \(\bar{Q}(i)\) are extracted.
    \item \textbf{Frequency domain (35 features):} To handle information density, we limit our frequency of interest range down to 0-1500 Hz and break it down to 5 equal intervals of 300 Hz from which we extract features 1 to 7 as \(F_{1-35}(x)\).
    \item \textbf{Time-frequency domain (28 features):} We used the \textit{coeif} wavelet to apply Discrete Wavelet Transform (DWT) and extracted features 1 to 7 from 3 levels of coefficients to \(TF_{1-28}(x)\) features. 
\end{itemize}

\bgroup
\def\arraystretch{1.5}
\begin{table*}
  \centering
  \footnotesize
  \begin{tabular}{|l|c|c|c|cccc|cccc|}
    \hline
    \multirow{2}{*}{Year}  &
    \multirow{2}{*}{Author} &
    \multirow{2}{*}{Input Features} &
    \multirow{2}{*}{Method} &
    \multicolumn{4}{c|}{Hold-out} & 
      \multicolumn{4}{c|}{LOOCV} \\
      &  &  &  & Acc & PR & SN & SP & Acc & PR & SN & SP \\
     \hline 
     \hline
    2015 & Zheng et al. & EMD & SVM & 0.5 & 0.44 & 0.66 & 0.38 & 0.49 & 0.49 & 0.61 & 0.37 \\
    \hline
    2016 & Nilanon et al. & Spectrograms & 2D-CNN & 0.65 & 0.75 & 0.52 & 0.80 & 0.55 & 0.54 & 0.49 & 0.62 \\
    \hline
    2017 & Li et al. & FFT & Logistic Regression & 0.58 & 0.64 & 0.52 & 0.66 & 0.54 & 0.50 & 0.49 & 0.57 \\
    \hline
    2017 & Boulares et al. & Mel Spectrogram & DNN + TL & 0.71 & 0.58 & 0.87 & 0.61 & 0.66 & 0.67 & 0.68 & 0.65 \\
    \hline
    \multirow{2}{*}{2018} & \multirow{2}{*}{Meintjes et al.} & \multirow{2}{*}{CWT} & KNN & 0.60 & 0.48 & 0.71 & 0.53 & 0.60 & 0.48 & 0.71 & 0.53 \\
     &  &  & SVM & 0.62 & 0.55 & 0.63 & 0.62 & 0.57 & 0.5 & 0.63 & 0.52 \\
    \hline
    2018 & Juniati et al. & DWT & KNN & 0.69 & 0.61 & 0.7 & 0.69 & 0.67 & 0.52 & 0.78 & 0.62 \\
    \hline
    2019 & Noman et al. & Signal + MFCC & CNN Ensemble & 0.58 & 0.64 & 0.52 & 0.66 & 0.54 & 0.50 & 0.49 & 0.57 \\
    \hline
    2019 & Nogueira et al. & MFCC & SVM & 0.65 & 0.75 & 0.52 & 0.80 & 0.55 & 0.54 & 0.49 & 0.62 \\
    \hline
    2019 & Shi et al. &  \cite{Springer} & AdaBoost & 0.82 & 0.80 & 0.86 & 0.79 & 0.81 & 0.77 & 0.88 & 0.75 \\
    \hline
    2020 & Khan et al. & MFCC & LSTM & 0.59 & 0.56 & 0.67 & 0.51 & 0.51 & 0.42 & 0.86 & 0.3 \\
    \hline
    2020 & Li et al. & Statistical & 1D-CNN & 0.61 & 0.57 & 0.61 & 0.62 & 0.63 & 0.51 & 0.78 & 0.54 \\
    \hline
    2020 & Alafif et al. & MFCC & 2D-CNN + TL & 0.68 & 0.56 & 0.79 & 0.62 & 0.64 & 0.61 & 0.69 & 0.6 \\
    \hline \hline
    2021 & \textbf{This work} & \textbf{Statistical + Sound} & \textbf{Ensemble} & \textbf{0.93} & \textbf{0.89} & \textbf{0.96} & \textbf{0.90} & \textbf{0.91} & \textbf{0.86} & \textbf{0.96} & \textbf{0.86} \\
    \hline
    
  \end{tabular}
  \caption{\label{tab:ourtabel_2}Fetus Gender Identification baseline results compared to our best approach. Results reported are Accuracy (Acc), Precision (PR), in addition to clinically-used Sensitivity (SN) and Specificity (SP) scores, with validation performed on a Hold-out set and leave-one-out cross-validation (LOOCV).}
\end{table*}
\egroup
\subsection{Fetal Gender Classification}
We propose an ensemble method of FGI using different ML algorithms. In an ensemble setting, many individually trained learning algorithms are teamed up to perform at the same classification/regression task that they were initially trained on together. Generally, in an ensemble method, there are three types of approaches that researchers have majorly taken. They have either used the Boosting or the Bagging method to combine homogeneous inefficient models or the Stacking method to increase the efficiency of heterogeneous weak learners. Same as the latter, we stack different individually-trained FGI classifiers and design a Meta-learner to learn from their mistakes. However, unlike common stacking methods where plenty of different classifiers are trained on the exact same feature representations, we train individual models using different feature representations of FPCG signals. These models are as follows:

\begin{enumerate}
    \item \textbf{Statistical Classifier:} We train an XGBoost Classifier \cite{Chen_2016} on the training set (\text{\footnotesize \(X\)}) to solely learn statistical features of denoised signals as \text{\footnotesize{\(C_{S}([T_{1-7}(X), F_{1-35}(X), TF_{1-28}(X)])\)}}.
    \item \textbf{Chroma Classifier:} Among sound features, this classifier learns to classify tonal features as \text{\footnotesize \(C_{R}(R(x))\)} using Linear discriminant analysis (LDA).
    \item \textbf{Mel Spectrogram Classifier:} We define an XGBoost Classifier as \text{\footnotesize \(C_{M}(M(x))\)} to classify denoised FPCG signals by their perceptual relevancy.
    \item \textbf{MFCC Classifier:} an SVM is trained to classify power spectrums as \text{\footnotesize \(C_{MF}(MF(x))\)}
    \item \textbf{CQT Classifier:} As our final individual learner, the \text{\footnotesize \(C_{Q}(Q(x))\)} classifier uses the K-Nearest Neighbors (KNN) algorithm to predict the gender using time-frequency information of a signal.
\end{enumerate}

Finally, having trained these models, we stack them to build an ensemble using the XGBoost Classifier as the meta-learner. Passing a denoised FPCG signal to each of the individual models, we obtain predicted class probability distributions which will be concatenated as an input vector to the Meta-learner, afterwhich the meta-learner would learn the correct final classification of fetal gender based on individually-trained models\textquotesingle{} class distributions.

\subsection{Performance Metrics}
To demonstrate our method's applicability and establish a baseline for our problem, we implemented methodologies that have either taken a classical ML or DL approach. We chose a diverse set of methodologies to compare various effects of preprocessing, feature extraction techniques, and classification methods. We report results using the two conventionally used ML classification metrics of Accuracy (Acc) and Precision (PR), in addition to clinically-used Sensitivity (SN) and Specificity (SP) scores. These metrics are once reported for each approach when it has been validated using a Hold-out set and again using Leave-One-Out Cross-Validation (LOOCV). It is noteworthy that at inference, we did not solely leave one sample out, since this may cause certain training samples belong to the subject (fetus) to which the test sample also belongs. To avoid this, we left all samples that belonged to a certain subject out (i.e. Leave-One-Subject-Out Cross-Validation) to evaluate a model. 

\section{Results \& Discussion} \label{sec:Results}
As shown in Table \ref{tab:ourtabel_1}, to form an ensemble, we compare the performance of different classifiers (i.e. KNN, SVM, XGBoost, LDA, and Logistic Regression (LR)) to choose the top-task-performing one as a final individual FGI model. For instance, discriminant MFCC features can be best identified using an SVM, whereas LR falls short. However, the impact of classifiers is shown to be less in an analogy to the denoising method. When trained on features extracted from the SCBSS-denoised signals, the classifiers are generally superior. The major difference between these neural denoising techniques is the structure of their network. Whereas the DAE is based on Autoencoders, the SCBSS relies on RNN and embeddings. Interestingly, where the first eliminates classification-worthy information through denoising, the latter filters such that models better understand them.  By stacking the chosen classifiers, we create an ensemble that performs better than any one model alone. 

Improvements made by our method are largely determined by two key factors. In contrast to most previously proposed ML methodologies (Table \ref{tab:ourtabel_2}), we let more than just one algorithm be in charge of classification. Studies such as \cite{8512284,Juniati}, who have also handcrafted their features of interest, mostly failed to accurately classify fetal gender correctly. This indicates that a combination of knowledge from individually studied sets of features can be beneficial in FGI.

The diversity of our features is another key to our methodology's performance.  We have experimentally gathered a set of discriminant features with various characteristics to generalize as much as possible using low-dimensional feature vectors. In contrast, DL models consist of complex architectures that demand training on a large amount of data. Moreover, DL approaches have focused on a particular feature representation \cite{7868810,Faiq,Mathiak} and have less addressed PCG classification using different representations. Therefore, it is unlikely that they can perform as well in FGI as in CVD detection. However, recent studies have tackled such situations by leveraging Transfer Learning \cite{9287514,9115633,Tseng2021CrossDomainTL}. The intuition here is that instead of extracting sound features of an audio signal and directly passing it as input to a neural network, the sound features are first visually represented and fed to a pretrained image network to obtain image features. In other words, instead of asking the model to learn discriminant sound features of two very similar classes from scarce data, the sound features are transformed into visual features where Transfer Learning can be applied. However, even these latest advancements came short at inference. We consider our extracted sets of statistical and sound features suitable for training efficient ML models to classify fetal gender using FPCGs. It is worth mentioning that to evaluate the true applicability of previously proposed DL methods on our dataset; we augmented our training set by using pitch shift or random noise and expanded our small dataset to the size of the dataset that an approach was initially tested on. Still, no significant changes were recorded after.

\section{Conclusion}
The prenatal identification of fetal gender is a day-to-day clinical task that is mostly carried out using Ultrasounds. In this work, we tackle the problem of fetus gender identification using acoustical sounds of the heart as recorded by PCG signals, and extend upon previous studies in CVD detection to the task presented in this paper. We established a baseline by experimenting with different denoising methods and feature extraction techniques to derive meaningful information from these signals to aid the modeling of fetus gender using classical ML and DL algorithms. Our proposed ensemble method significantly improved the baseline by correctly classifying 91\% of the FPCG signals among unseen subjects. Our results motivate further work in robust denoising techniques, and the development of end-to-end systems for modeling heart activity and FGI.


\bibliography{aaai22}
\bibliographystyle{aaai}

\end{document}